\documentclass[aps,preprint]{revtex4}%
\usepackage{amsfonts}
\usepackage{amsmath}
\usepackage{amssymb}
\usepackage{graphicx}%
\begin{document}
\title[Short title for running header]{On the Origin of the Tunneling Asymmetry in the Cuprate Superconductors}
\author{Tao Li, Fan Yang and Yong-Jin Jiang}
\affiliation{Center for Advanced Study, Tsinghua University, Beijing 100084, P.R.China}

\begin{abstract}
We argue the coherent part of the spectral weight always contribute
symmetrically to the STM spectrum at sufficiently low energy and the tunneling
asymmetry is a manifestation of the incoherent part of the electron spectrum.
By subtracting the particle side spectrum from the hole side spectrum in the
published STM data of Pan \textit{et al} on $BSCCO_{2212}(T_{c}=84K)$, we find
the difference spectrum show a well defined gap structure at 25 meV. We argue
this gap may represent a new energy scale of the system, namely the energy
scale for spin-charge recombination in this system.

\end{abstract}
\volumeyear{year}
\volumenumber{number}
\issuenumber{number}
\eid{identifier}
\startpage{1}
\endpage{10}
\maketitle

\bigskip The scanning tunneling microscopy(STM) plays an important role in the
study of the high temperature superconductors since it can provide local
information on the single particle properties with ultrahigh energy
resolution. A striking feature in the STM spectrum of the high temperature
superconductors is their remarkable particle-hole asymmetry. The hole side of
the spectrum always dominate the particle side of the spectrum in hole doped
cuprates\cite{1}.

\bigskip This asymmetry is not at all surprising if we take the high
temperature superconductors as doped Mott insulators described by the $t-J$
model. In such a doped Mott insulator, an added electron has a reduced
probability to contribute to the spectral weight in the low energy subspace of
no doubly occupied site. More specifically, if the hole density in the system
is $x$, then the total spectral weight in the particle side of the spectrum
will be reduced to $x,$ while the total spectral weight in the hole side of
the spectrum is not affected by the no double occupancy constraint. Thus the
total spectral weight is particle-hole asymmetric for small $x$. However, such
an asymmetry on the total spectral weight tell us nothing about the
distribution of the spectral weight at low energy. To address the problem of
tunneling asymmetry near the chemical potential, we need more detailed
information on the low energy excitation spectrum of the system.

Rantner and Wen addressed this problem in the slave-Boson mean field theory of
the $t-J$ model\cite{2}. In their theory, the electron spectrum in the
superconducting state can be divided into the coherent part and the incoherent
part. The particle-hole asymmetry of the tunneling spectrum comes from the
incoherent part of the spectral weight which is totally absent in the particle
side of spectrum at zero temperature. In the slave-Boson mean field theory,
the electron is split into two parts, $c_{i\sigma}=b_{i}^{\dagger}f_{i\sigma}%
$, in which $b$ is the operator for Bosonic holon and $f$ is the operator for
Fermionic spinon. In the mean field theory, the superconductivity is achieved
by pairing up the spin one-half spinon and Bose condense the charge one holon.
The electron propagator is the convolution of the spinon propagator and the
holon propagator. In the presence of the holon condensate, the electron
spectrum can be divided into the coherent part and the incoherent part. The
coherent part of spectrum originates from the holon condensate and has well
defined dispersion(which is nothing but the spinon dispersion). The incoherent
part of the spectral weight originates from holon excitation above the
condensate and constitute a continuum in the electron spectrum. In the mean
field theory, the incoherent part of the spectral weight is totally absent in
the particle side of the spectrum at zero temperature since all holon are
condensed into the lowest energy state while by adding an electron into system
we must remove a holon from the system. Thus if we neglect the asymmetry of
spinon dispersion near the chemical potential, the tunneling asymmetry can be
totally attributed to the incoherent part of spectral weight of electron propagator.

Recently, Anderson and Ong addressed the same problem with a variational
approach\cite{3}. They constructed explicitly the variational wave function
for the ground state and the excited state of the $t-J$ model following the
original RVB idea. In their treatment, the particle-hole asymmetry is taken
into account explicitly in the variational wave function by the introduction
of a so called fugacity factor. This fugacity factor plays part of the role of
the Gutzwiller projection into the subspace of no double occupancy. In this
theory, most of the electron spectral weight is coherent(has well defined
dispersion and corresponds to well defined wave function). Unlike Rantner's
theory, the tunneling asymmetry is not related to the incoherent part of the
spectral weight.

\bigskip In this paper, we argue the coherent part of the spectral weight of
an electronic system always contribute symmetrically to the STM spectrum at
sufficiently low energy and the particle-hole asymmetry at low energy is a
manifestation of the incoherent part of the electron spectrum. By subtracting
the particle side spectrum from the hole side spectrum in published STM data
of Pan \textit{et al }(on $BSCCO_{2212}$ with $T_{c}=84K$), we find the
tunneling asymmetry appear only above a well defined energy gap of about 25
meV. We argue this gap may represent a new energy scale of the system, namely
the energy scale for spin-charge recombination in this system.

By definition, the coherent part of the spectral weight of an electron(or a
hole) comes from the contribution of quasiparticle(or quasihole). In the
Landau theory of Fermi liquid, the quasiparticle(quasihole) plays a dual role.
On the one hand, the quasiparticle(quasihole) can be thought of as a
particle-like(hole-like) elementary excitation on the ground state of a $N$
particle system. On the other hand, the quasiparticle(quasihole) can also be
thought of as a constituent of the grounds state of the $N+1$ particle($N-1$
particle) system, provide that the quasiparticle(quasihole) is on the Fermi
surface. The coherent part of the spectral weight for adding an electron into
the system on the Fermi surface is thus equal to the square of the matrix
element of the electron creation operator between the ground state of $N$
particle system and the ground state of $N+1$ particle system,%

\[
Z_{N}^{a}=\left\vert \left\langle g_{_{N+1}}\right\vert c_{_{k}}^{\dagger
}\left\vert g_{_{N}}\right\rangle \right\vert ^{2}%
\]

while the coherent part of the spectral weight for removing an electron from
the system on the Fermi surface is equal to the square of the matrix element
of electron annihilation operator between the ground state of $N$ particle
system and the ground state of $N-1$ particle system,%

\[
Z_{N}^{r}=\left\vert \left\langle g_{_{N-1}}\right\vert c_{_{k}}\left\vert
g_{_{N}}\right\rangle \right\vert ^{2}=Z_{N-1}^{a}%
\]
In the thermodynamic limit, we have%

\[
Z_{N}^{r}=Z_{N-1}^{a}\simeq Z_{N}^{a}%
\]

thus the coherent part of the spectral weight is particle-hole symmetric. This
conclusion is consistent with that of the slave-Boson mean field theory. In
the presence of the superconducting pairing, the above argument is not
applicable. However, since the no double occupancy constraint(which is
believed to be the ultimate reason for particle-hole asymmetry in this system)
occurs at an energy scale much higher than that of the superconducting paring,
we do not expect the latter to change our conclusion essentially.

Since the coherent part of the spectral weight is particle-hole symmetric, the
tunneling asymmetry must be attributed to the incoherent part of the electron
spectrum. Thus from the STM spectrum we can extract information on the
incoherent part of the electron spectrum. For a strongly correlated system
like cuprates, such information is of great importance. From such information
we can in principle figure out the mechanism by which a bare particle decay
into the many particle excitations and thus on the nature of the many particle
excitations itself. In the context of the cuprates, two mechanisms of
generating electron incoherence are frequently discussed. The first is by
scattering with some bosonic collective mode in the system. Among such
collective modes, the neutron resonance mode in the spin channel\cite{4,5} and
the oxygen stretching phonon mode are most frequently involved\cite{6}. Such a
mechanism can provide a natural energy scale for the incoherent part of the
spectral weight if the collective mode itself is gapped. The second mechanism
to generate electron incoherence is by spin-charge separation as we have
discussed in the slave-Boson mean field theory\cite{2}. In this mechanism, the
electron split into two elementary excitations carrying its spin and charge
quantum number separately. The independent propagation of the two elementary
excitations lead to electron incoherence. We note such a mechanism do not
provide a natural energy scale for the incoherent part of the spectral weight.

Along this line of reasoning, we have extracted the asymmetric part of the
STM\ spectrum by subtracting the particle side spectrum from the hole side
spectrum in the published STM data of Pan \textit{et al }on $BSCCO_{2212}%
$\cite{1}. The result is shown in Figure 1 and is rather striking. We find the
tunneling asymmetry occurs only above a well defined gap about 25 meV. This
indicate that the incoherent part of the electron spectral weight has a well
defined energy scale. This is the main result of this paper.
%TCIMACRO{\FRAME{ftbpF}{3.0519in}{1.946in}{0pt}{}{}{fig1a.eps}%
%{\special{ language "Scientific Word";  type "GRAPHIC";  display "USEDEF";
%valid_file "F";  width 3.0519in;  height 1.946in;  depth 0pt;
%original-width 6.665in;  original-height 5.139in;  cropleft "0";
%croptop "1";  cropright "1.2382";  cropbottom "0";
%filename 'fig1a.EPS';file-properties "XNPEU";}}}%
%BeginExpansion
\begin{figure}
[ptb]
\begin{center}
\includegraphics[
trim=0.000000in 0.000000in -1.587603in 0.000000in,
natheight=5.139000in,
natwidth=6.665000in,
height=1.946in,
width=3.0519in
]%
{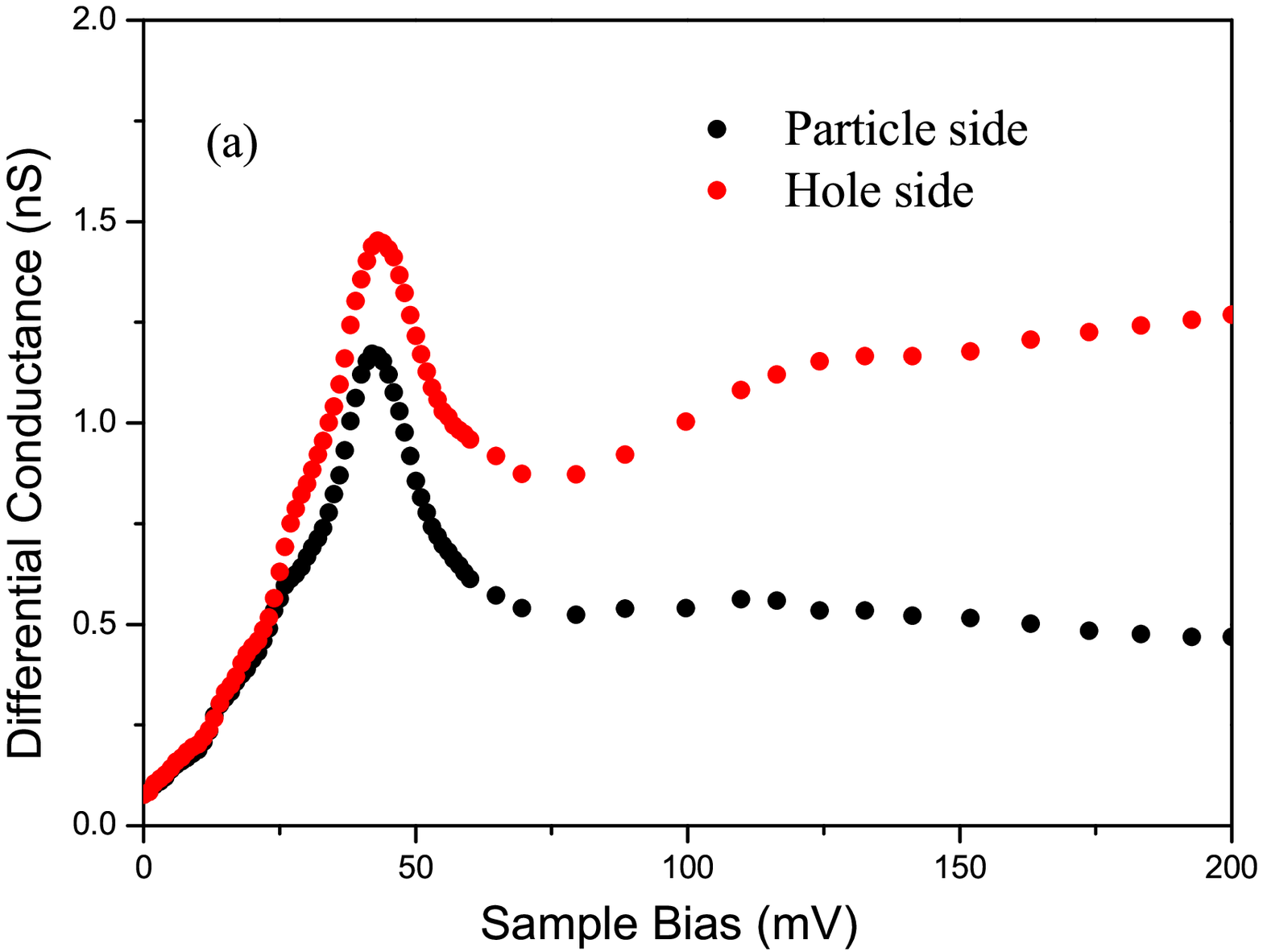}%
\end{center}
\end{figure}
%EndExpansion
%

%TCIMACRO{\FRAME{ftbpFU}{3.0128in}{1.8173in}{0pt}{\Qcb{The STM\ spectrum of
%$BSCCO_{2212}$($T_{c}=84K$). (a) Original data of Pan \textit{et al }%
%\cite{1}(data points above 60 mV are only partly reproduced in this figure for
%clearness). (b)The difference spectrum obtained by subtracting the particle
%side spectrum from the hole side spectrum. The inset show the difference
%spectrum in the whole range of energy measured by Pan \textit{et al}. }}%
%{}{fig1b.eps}{\special{ language "Scientific Word";  type "GRAPHIC";
%display "USEDEF";  valid_file "F";  width 3.0128in;  height 1.8173in;
%depth 0pt;  original-width 6.665in;  original-height 5.1656in;  cropleft "0";
%croptop "1";  cropright "1.1688";  cropbottom "0";
%filename 'fig1b.EPS';file-properties "XNPEU";}}}%
%BeginExpansion
\begin{figure}
[ptb]
\begin{center}
\includegraphics[
trim=0.000000in 0.000000in -1.125052in 0.000000in,
natheight=5.165600in,
natwidth=6.665000in,
height=1.8173in,
width=3.0128in
]%
{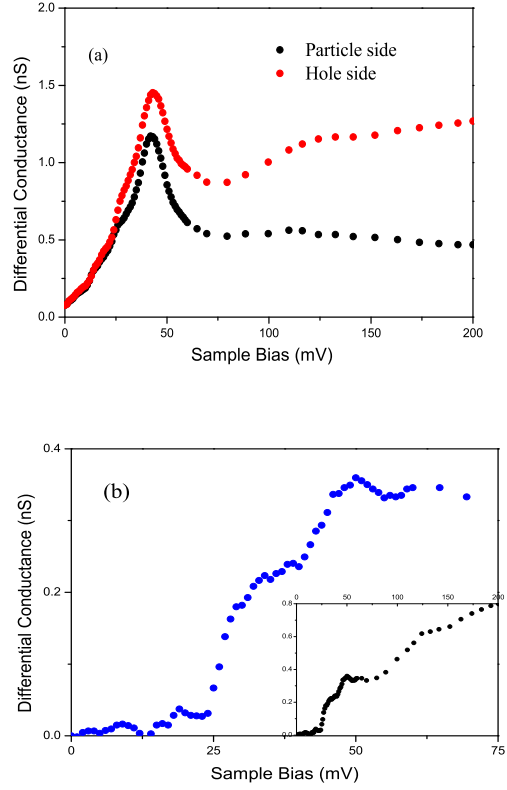}%
\caption{The STM\ spectrum of $BSCCO_{2212}$($T_{c}=84K$). (a) Original data
of Pan \textit{et al }\cite{1}(data points above 60 mV are only partly
reproduced in this figure for clearness). (b)The difference spectrum obtained
by subtracting the particle side spectrum from the hole side spectrum. The
inset show the difference spectrum in the whole range of energy measured by
Pan \textit{et al}. }%
\end{center}
\end{figure}
%EndExpansion

Now we discuss the possible origin of the energy gap in the difference
spectrum. In the collective mode scattering scenario, the incoherent part of
the spectral weight do show a energy scale if the collective mode itself is
gapped. However, the extracted gap for the incoherent part of the spectral
weight, namely, 25meV, seems too small for such a explanation to apply. In the
system studied, the energy of the neutron resonance mode $E_{r}$ can be
estimated from the $T_{c}-E_{r}$ scaling to be $E_{r}\sim37$ meV. Thus we
expect the incoherent part of the spectral weight to begin above 37 meV. The
phonon mechanism involve an energy scale even higher than that of the neutron
mode (70 meV) and is thus also too high to explain the gap in the difference
spectrum. At the same time, this mechanism provide no simple understanding of
the dominance of the hole side spectrum over the particle side spectrum. For
these reasons, the collective mode scattering is not likely the main origin of
the electron incoherence and thus the tunneling asymmetry of the cuprates.
However, we think the scattering with the neutron resonance mode may be
responsible for the broad peak centered at 50 meV in the difference spectrum.

\bigskip Now we analyse the tunneling asymmetry in the spin-charge separation
scenario. In this scenario, the dominance of the hole side spectrum is
directly related to the no double occupancy constraint, as we have discussed
in the slave-Boson mean field theory of the $t-J$ model. One problem with this
scenario is that the spin-charge separation itself do not provide a natural
energy scale for the incoherent part of the spectral weight. In the slave
Boson mean filed theory, the incoherent part of the spectral weight start at
zero energy since the density of state for a holon dispersion is a constant at
low energy in the two dimensional system. Here we propose that the energy gap
in the difference spectrum may represent a new energy scale in the system,
namely, the energy scale for spin-charge recombination. Below this energy
scale, the spectral weight is totally coherent and the electron propagate as
an integrated part in the system. Above this energy scale, an electron
dissociated into two parts carrying its spin and charge quantum number
separately. The two parts propagate independently and make up the incoherent
part of the electron spectrum. The detail of the spin-charge recombination
process is still unknown and we call for further theoretical study on this problem.

\ \ The author would like to thank members of the HTS group at CASTU for
discussion. T. Li is supported by NSFC Grant No. 90303009.

\end{document}